\documentclass[aps,pra,reprint,twocolumn,floatfix,superscriptaddress,longbibliography]{revtex4-1}
\usepackage{amsfonts, amsmath, amssymb}
\usepackage{graphicx}
\usepackage{dcolumn}
\usepackage{bm}
\usepackage{bbm}
\usepackage{color}
\usepackage[utf8]{inputenc}
\usepackage[colorlinks,citecolor=blue,linkcolor=blue,urlcolor=blue]{hyperref}
\usepackage{soul}

\usepackage{color}
\usepackage{xcolor}
\usepackage{siunitx}

\newcommand{\new}[1]{{\color{black} #1}}

\newcommand{\micron}{\rm{\si{\micro\meter}}}
\newcommand{\fcal}{\mathcal{F}}
\newcommand{\be}{\begin{equation}}
\newcommand{\ee}{\end{equation}}
\newcommand{\ba}{\begin{eqnarray}}
\newcommand{\ea}{\end{eqnarray}}
\newcommand{\bea}{\begin{equation}\begin{aligned}}
\newcommand{\eea}{\end{aligned}\end{equation}}

\newcommand{\abs}[1]{\left\lvert #1 \right\rvert}

\def\rmi{{\rm {i}}}
\def\d{{\rm {d}}}
\def\tr{{\rm{Tr}}}
\newcommand{\rhohat}{\hat{\rho}}

\newcommand{\psilp}{\psi}
\newcommand{\olp}{\omega_{\rm{LP}}^{k=0}}

\newcommand{\Dlp}{\Delta}
\newcommand{\mlp}{m}
\newcommand{\glp}{g}
\newcommand{\gamlp}{\gamma}
\renewcommand{\vec}{\mathbf}
\include{physics}

\newcommand{\id}{\mathbbm{1}}

\newcommand{\psiss}{\psi_{\rm{SS}}}
\newcommand{\psid}{n_D}
\newcommand{\psibar}{n_D^\mathrm{SS}}
\newcommand{\lcal}{\mathcal{L}}

  {\left\lbrace\begin{array}{@{}l@{}}}%
  {\end{array}\right.}

\usepackage{lipsum}

\begin{document}
\title{Dissipative phase transition with driving-controlled spatial dimension \\ and diffusive boundary conditions \\ }

\author{Zejian Li}
\thanks{These two authors contributed equally}
\affiliation{Laboratoire Mat\'{e}riaux et Ph\'{e}nom\`{e}nes Quantiques (MPQ), Universit\'{e} de Paris, CNRS-UMR7162, Paris 75013, France}

\author{Ferdinand Claude}
\thanks{These two authors contributed equally}
\affiliation{Laboratoire Kastler Brossel, Sorbonne Universit\'{e}, CNRS, ENS-Universit\'{e} PSL, Coll\`{e}ge de France, Paris 75005, France}

\author{Thomas Boulier}
\affiliation{Laboratoire Kastler Brossel, Sorbonne Universit\'{e}, CNRS, ENS
-
Universit\'{e} PSL, Coll\`{e}ge de France, Paris 75005, France}

\author{Elisabeth Giacobino}
\affiliation{Laboratoire Kastler Brossel, Sorbonne Universit\'{e}, CNRS, ENS
-
Universit\'{e} PSL, Coll\`{e}ge de France, Paris 75005, France}

\author{Quentin Glorieux}
\affiliation{Laboratoire Kastler Brossel, Sorbonne Universit\'{e}, CNRS, ENS
-
Universit\'{e} PSL, Coll\`{e}ge de France, Paris 75005, France}

\author{Alberto Bramati}
\affiliation{Laboratoire Kastler Brossel, Sorbonne Universit\'{e}, CNRS, ENS
-
Universit\'{e} PSL, Coll\`{e}ge de France, Paris 75005, France}

\author{Cristiano Ciuti}
\affiliation{Laboratoire Mat\'{e}riaux et Ph\'{e}nom\`{e}nes Quantiques (MPQ), Universit\'{e} de Paris, CNRS-UMR7162, Paris 75013, France}

\begin{abstract}
We investigate theoretically and experimentally a first-order dissipative phase transition, with diffusive boundary conditions and the ability to tune the spatial dimension of the system. 
 The considered physical system is a planar semiconductor microcavity in the strong light-matter coupling regime, where polariton excitations are injected by a quasi-resonant optical driving field. 
The spatial dimension of the system from 1D to 2D is tuned by designing the intensity profile of the driving field. We investigate the emergence of criticality by increasing the spatial size of the driven region.
 The system is nonlinear due to polariton-polariton interactions and the boundary conditions are diffusive because the polaritons can freely diffuse out of the driven region.
We show that no phase transition occurs using a 1D driving geometry, while for a 2D geometry we do observe both in theory and experiments the emergence of a first-order phase transition. 
The demonstrated technique allows all-optical and in-situ control of the system geometry, providing a versatile plateform for exploring the many-body physics of photons.
\end{abstract}

\date{\today}
\maketitle

{\it Introduction.---}
The study of phase transitions and critical phenomena is at the heart of condensed matter physics and material science \cite{stanley_introduction_1987}. In classical systems, thermal phase transitions, such as that from a liquid to a solid phase, occur at finite temperature and are driven by thermal fluctuations. In a closed quantum system, phase transitions can happen at zero temperature, where the system is in its ground state, driven by quantum fluctuations due to the competition of non-commuting terms in the Hamiltonian \cite{Sachdev2009}. On the other hand, open quantum systems subject to driving and dissipation, can exhibit dissipative phase transitions for the non-equilibrium steady state,  where the physics is decided by the rich interplay between the Hamiltonian evolution, dissipation-induced fluctuations and driving.

Driven-dissipative phase transitions have been theoretically studied for various systems, such as photonic resonators \cite{Bartolo2016,Carmichael2015,Weimer2015,Benito2016,MendozaArenas2016,Casteels2016,Casteels2017,Casteels2017Quantum,FossFeig2017,Biondi2017,Biella2017Phase,Savona2017Spontaneous,Verstraelen2020,Vicentini2018Critical}, exciton-polariton condensates \cite{Sieberer2013dynamical,Sieberer2014Nonequilibrium,Altman2015Two,Dagvadorj2021First}, and spin systems \cite{Lee2013Unconventional,Jin2016Cluster,Kessler2012Dissipative,Lee2011Antiferromagnetic,Chan2015Limit,Rota2017Critical,Overbeck2017Multicritical,Roscher2018phenomenology}. Experimental investigations have studied dissipative phase transitions in single-mode semiconductor microcavity pillars \cite{Rodriguez2017Probing} and 
superconducting  resonators \cite{Fink2017Signatures, Fitzpatrick2017Observation}. Recent theoretical works \cite{Vicentini2018Critical,FossFeig2017} predicted that in a driven-dissipative lattice of photonic resonators with Kerr nonlinearities a first-order dissipative phase transition emerges in two-dimensional (2D) lattices (with periodic boundary conditions), while in 1D chains there is no critical phenomenon. \new{Note that in general the emergence of a phase transition can be drastically affected by its spatial dimensionality \cite{Sachdev2009}.}

In this work, we explore both theoretically and experimentally the role of spatial dimension for a dissipative phase transition using a planar semiconductor microcavity, where polariton excitations are injected via quasi-resonant driving. 
We propose theoretically and implement experimentally an all-optical way to enforce the dimensionality via the spatial shape of the driving beam. In particular, we consider a top-hat spot with constant driving intensity. The shape of the spot can be tailored in-situ to create a 2D or 1D geometry \new{\footnote{\new{In principle, one could also use etching techniques, that may lead to better defined structures, to study the effect of spatial dimension on dissipative phase transitions. However, the present all-optical approach provides  flexibility and can be used to explore the effect of a gradual change of dimensionality {\it in situ} using the same sample.}}}. This scheme also features ``diffusive" boundary conditions, since the polaritons can diffuse away from the driven region. While increasing the spatial size of the spot, which is the thermodynamic limit in the present context, we show that a first-order phase transition occurs using a 2D geometry, while it disappears in the 1D configuration, providing a first experimental demonstration of the role of dimensionality in driven-dissipative phase transitions of photonic systems.

{\it Theoretical model.---}
\begin{figure*}[htbp]
    \centering
    \includegraphics[width=1\linewidth]{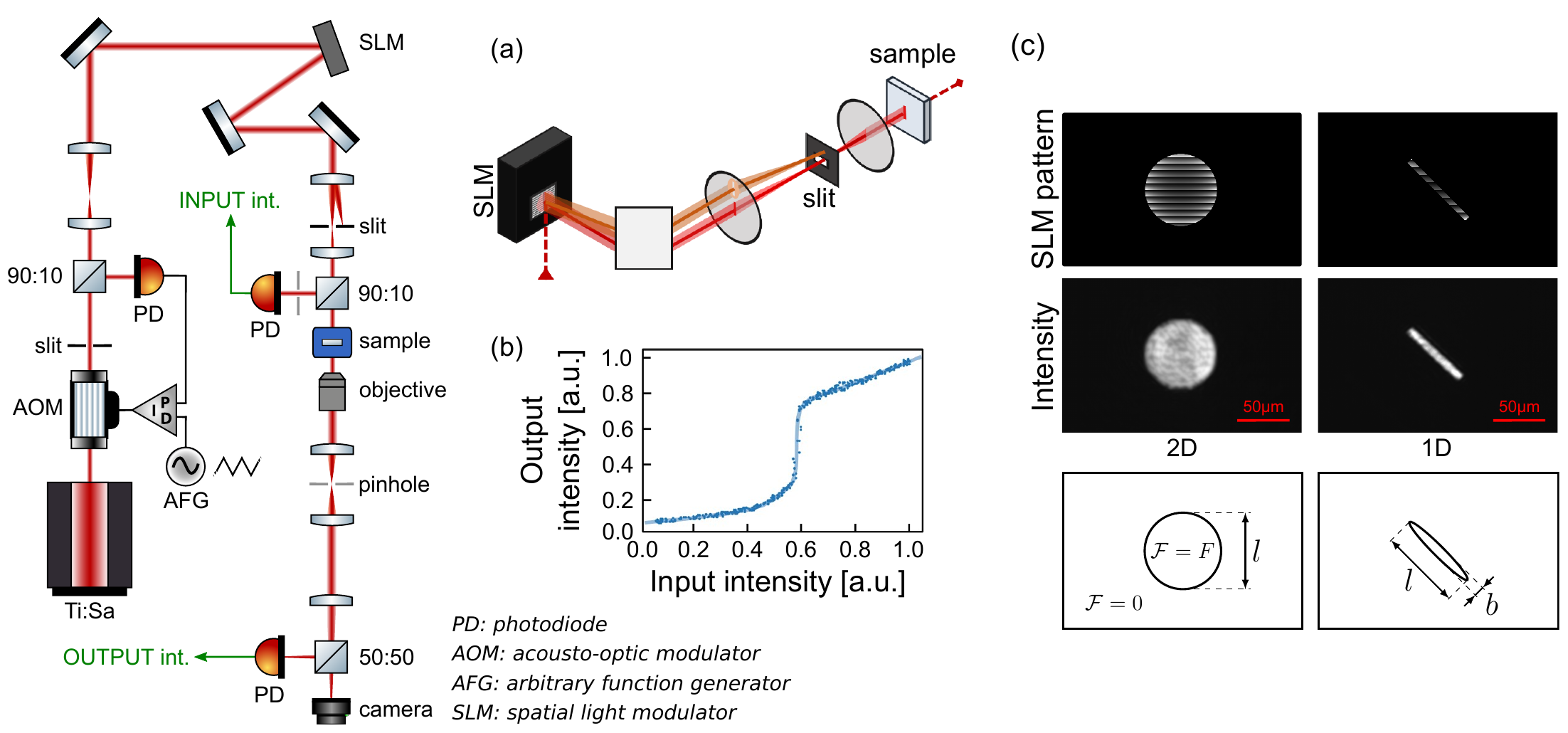}
    \caption{Sketch of the experimental setup. The laser is slaved using a proportional–integral–derivative (PID) controller, an arbitrary function generator (AFG) and an acousto-optic modulator (AOM) loop to produce a power ramp; its intensity profile is reshaped using a spatial light modulator (SLM). Two photodiodes (PD) measure the power inside disks of diameter $l_D = 5\micron$ at the center of the beams at the sample input and output. (a) Pump intensity profile shaping method: the light (dark) beam represents the zero (first) order of the diffracted beam from the SLM. (b) Output intensity from the sample as a function of the input intensity, plotted for a pump detuning of $\Dlp=\gamma$ and a 2D top-hat drive of diameter $l=30\micron$.  (c) SLM phase pattern (upper) for obtaining 2D (left) and 1D (right) flat-top beam profiles (middle) of different sizes and intensities (bottom).}
    \label{fig:setup}
\end{figure*}
Consider a planar semiconductor microcavity in the strong light-matter coupling regime, where polariton excitations are coherently injected by a quasi-resonant optical drive. The system dynamics can be described in terms of the lower polariton field $\hat{\psilp}(\vec{r},t)$ \cite{Carusotto2013Quantum}, where $\vec{r}=(x,y)$ are in-plane coordinates parallel to the cavity mirrors.
Within the mean-field approximation \footnote{See Supplementary material for a derivation of the mean-field dynamics from the quantum master equation \cite{Breuer2007}, and for a benchmark against the truncated Wigner method \cite{Vogel1989,Carmichael1999Statistical}, as a justification of this approximation.}, the time evolution of the mean-field $\psilp(\vec{r},t)=\langle \hat{\psi}(\vec{r},t) \rangle$ in the frame rotating at the driving frequency $\omega_d$ can be described by the dynamical equation \cite{Carusotto2013Quantum}: 
\bea\label{eq:gp}
\rmi \dfrac{\partial}{\partial t}\psilp(\vec{r},t) =& \left( -\Dlp - \dfrac{\hbar}{2\mlp}\nabla^2 \right)\psilp(\vec{r},t) \\&+ \glp\abs{\psilp(\vec{r},t)}^2\psilp(\vec{r},t) - \rmi\dfrac{\gamlp}{2}\psilp(\vec{r},t) \\&+ \fcal(\vec{r}),
\eea
where $\hbar$ is the Planck constant, $\Dlp = \omega_d - \olp$ is the detuning of the drive with respect to the $k=0$ mode of the lower polariton branch, $\mlp$ is the lower polariton effective mass, $\glp$ is the polariton-polariton interaction constant, $\gamlp$ is the lower polariton loss rate and $\mathcal{F}(\vec{r})$ encodes the amplitude and spatial shape of the coherent drive.

In the following, We adopt a top-hat driving scheme [see Fig. \ref{fig:setup} (c)], where the amplitude $\fcal(\vec{r})$ is defined by
\bea
    \fcal(\vec{r}) = F \id_A(\vec{r}),
\eea
where $\id_A$ is the indicator function of a compact region $A$ of the plane, such that the drive is constant within the region $A$ and zero elsewhere. To force a 1D geometry, the driving region will be chosen as an elliptical spot with fixed minor axis $b$ and variable major axis $l \gg b$. To induce a 2D geometry, instead, the driving region will be chosen as a circular disk of variable diameter $l$. 
% \bea
%     A_{1\rmD}(l) = \left\{ (x,y) \mid \left( 4x/l \right)^2 + \left( 4y/b \right)^2 \leq 1 \right\},
% \eea
% which is an elliptical spot with fixed minor axis $b$ and  variable major axis $l\gg b$. To induce a 2D geometry, instead, the driving will be chosen as
% \bea
%     A_{2\rmD}(l) = \left\{ (x,y) \mid \left( 4x/l \right)^2 + \left( 4y / l \right)^2 \leq 1 \right\},
% \eea
% which is a disk of variable diameter $l$. 
Note that the only difference between the 1D and 2D configurations is the spatial shape of the top-hat drive \footnote{Note that the way we distinguish 1D and 2D is not via the absolute size of the top-hat spot, but the different ways they approach the thermodynamic limit: in 1D only the major axis $l$ increases with $b$ fixed, such that in the thermodynamic limit $\lim_{l\rightarrow\infty}b/l=0$, whereas in 2D both axes increase at the same rate, keeping the spot always circular and its aspect ratio constant}, while the planar microcavity sample is the same. The boundary conditions in terms of the driven region are therefore of diffusive nature, which means that the polaritons can freely diffuse and decay out of the driving spot.

In order to probe a dissipative phase transition with respect to the driving intensity $I=\abs{F}^2$, we will be interested in the steady-state polariton density averaged over a disk $D$ of diameter $l_D$ at the center of the driven region:
% \bea
%     \psibar=& \dfrac{1}{\mu(D)}\int_D\d^2\vec{r} \langle\hat{\psi}^\dagger_\mathrm{SS}(\vec{r})\hat{\psi}_\mathrm{SS}(\vec{r})\rangle\\ \simeq &\dfrac{1}{\mu(D)}\int_D \d^2\vec{r} \abs{\psiss(\vec{r})}^2,
% \eea
\bea
    \psibar=&\dfrac{1}{\mu(D)}\int_D \d^2\vec{r} \abs{\psiss(\vec{r})}^2,
\eea
where $\mu(D)$ denotes the area of the disk $D$ and $\psiss$ is the steady state field such that $\partial_t\psiss=0$.
% ; note that the second equality comes from the mean-field approximation. 
In the thermodynamic limit of $l\rightarrow\infty$, a transition between two phases is characterized by the non-analytical behavior of $\psibar$ when $I$ tends to some critical value $I_c$. Formally, a transition of order $M$ can be described as \cite{Minganti2018Spectral} 
\bea\label{eq:ordm}
    \lim_{I\rightarrow I_c}\left\lvert \dfrac{\partial^M}{\partial I^M}\lim_{l\rightarrow\infty}\psibar \right\rvert = +\infty.
\eea

In this letter we will present a first order ($M=1$) phase transition, that is a discontinuity of steady-state polariton density $\psibar$ with respect to the drive intensity $I$, which are the two quantities that we measure in our experiments.
% As we will show in the results, the behavior of the slope in 2D indicates the divergence of the first-derivative ($M=1$) in the thermodynamic limit, therefore the emergence of a first-order phase transition.

{\it Experimental setup.---}
The sample used in our experiments is a 2$\lambda$ GaAs high-finesse semiconductor microcavity cooled to the temperature of 4K in an open-flow helium cryostat. The cavity embeds three In$_{0.04}$Ga$_{0.96}$As quantum wells (QWs) between a pair of distributed Bragg mirrors made of 21 (top) and 24 (bottom)
alternated layers of GaAs/AlAs. Each QW is located on an antinode of the cavity electromagnetic field to have a strong coupling of QW excitons to the cavity photons, giving rise to the exciton-polariton modes. The cavity spacer has a small wedge ($\hbar w\simeq$ 0.7 µeV/µm) whereby the photon-exciton detuning can be finely adjusted to around 0 meV by changing the excitation position. At this detuning the lower polariton branch has an effective mass $\mlp$ = $5.7\times 10^{-5} m_\mathrm{e}$, where $m_\mathrm{e}$ is the bare electron mass.
The Rabi frequency, the lower-polariton decay rate and the polariton-polariton interaction constant are respectively measured to be $\hbar \Omega_R = 5.1$ meV, $\hbar\gamlp = 0.08$ meV, and $\hbar \glp = 0.01$ meV$\cdot\micron^2$ \footnote{As we will choose a detuning of $\Dlp=\gamlp$ with respect to the lower polariton branch in the next section, this means that the Rabi frequency $\hbar\Omega_R$, which determines the minimum splitting between the lower and upper polariton branches, is much larger than all the other energy scales in the problem. We can therefore consider effectively only the lower polariton in our theoretical treatment \cite{Carusotto2013Quantum}. }.

The polaritons are excited by a circularly polarized continuous-wave Ti:Sapphire laser whose output Gaussian mode is reshaped with a spatial light modulator (SLM) (Fig. \ref{fig:setup}). The SLM liquid crystal matrix plane is imaged on that of the cavity and contains a blazed grating of tunable contrast, which diffracts in the first order a fraction of the driving field intensity. The first order component is sent at normal incidence  through the cavity %so that the polariton fluid does not have any imposed flow
, while the non-diffracted part (zero order) is blocked in the Fourier plane with a slit [Fig. \ref{fig:setup}(a)].

By locally adjusting the grating contrast,  we modify the intensity distribution between the zero and the first orders. 
In this way, with a well-calibrated anti-Gaussian contrast gradient - minimum at the center and maximum at the edge of the spot - we produce in the first order a flat top-hat intensity profile.
Then, by adding a non-diffracting mask over the grating, one can select which area of the beam profile is reflected into the first order. 
Thus, the shape of the driving spot in the cavity reproduces the one defined by the contours of the mask [Fig. \ref{fig:setup}(c)]. 
With this reshaping method, we can go from a 2D circular driving spot to a 1D elliptical one,  by configuring the SLM respectively with a blazed grating masked by a circular aperture or by a narrow slit (see Fig. \ref{fig:setup}). In the following, the spot sizes in the cavity plane are tuned by changing the mask dimension. For the 1D geometry, the minor axis of intensity profile is set at $b = 6.4\micron$ \new{\footnote{\new{This value is chosen such that it is large compared to the optical wavelength of the laser to avoid undesirable diffraction effects so as to produce a well-defined top-hat. At the same time, it should be small enough to ensure that the crossover slope for the smallest top-hat is mild enough to be measured experimentally, which allows us to study the asymptotic behavior (convergence or divergence) of the growing slope.}}}

In order to probe the phase transition, an acousto-optic modulator (AOM) modulates the driving field power with a low-frequency ramp (200 Hz) of sufficient amplitude to be able to scan a wide range of polariton density.
The input and output intensities of the cavity are measured using two photodiodes which detect through pinholes (of diameter $l_D = 5 \micron$ \new{\footnote{\new{Note that as the probing disk $D$ is placed concentrically with the top-hat profile, the chosen value ensures that it is always contained within the driven region. However, we expect that asymptotically (in the limit of large $l$), the observed effects should not depend on its specific position as long as the probing disk is far enough from the boundary (or the edges of the major axis in the 1D case) of the top-hat.}}}
% \cite{Note999}
) at the center of the driving spot. Thus, the polariton density is directly observed as a function of the driving intensity by plotting, one with respect to the other, the powers detected by the two photodiodes [see Fig. \ref{fig:setup}(b)].

{\it Results and discussion.---}
\begin{figure*}[t]
\centering
    \includegraphics[width=0.87\linewidth]{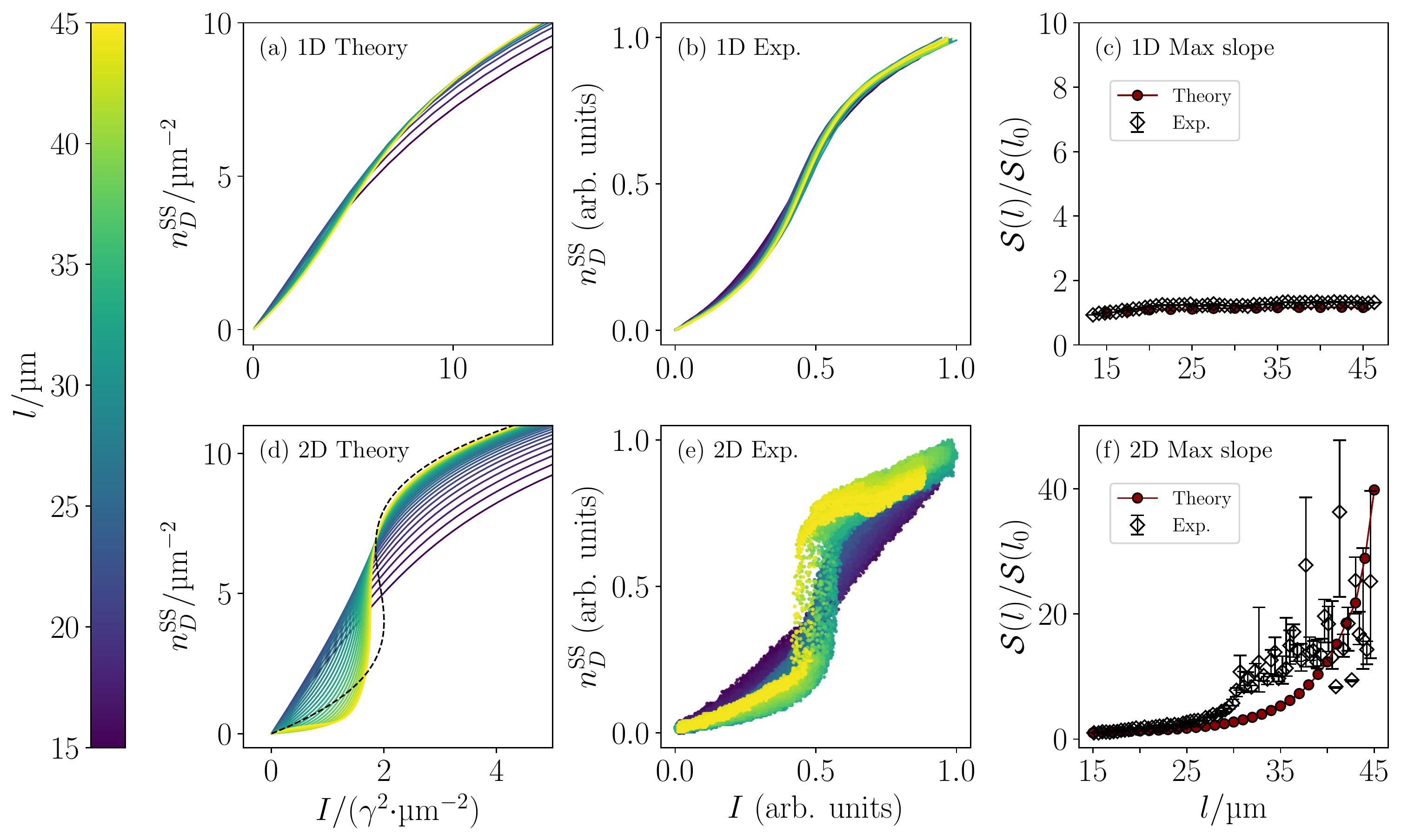}
    \caption{(a) [(b)] Theoretical (experimental) results for the steady-state polariton density $\psibar$ averaged over the probing disk as a function of the drive intensity $I$ for different top-hat spot sizes $l$ (see colorbar) in the 1D configuration with detuning $\Dlp = \gamlp$. (c) The maximum derivative $\mathcal{S}(l)$ for each top-hat size $l$ normalized by the maximum derivative at $l_0=15\micron$, for both theoretical and experimental results (see legend). (d)-(f) The same quantities as in (a)-(c) for the 2D configuration. The dashed line in (\new{d}) is the prediction of the mean-field theory in Eq.  (\ref{eq:MFbis}). Note that as the top-hat increases in size, the slope in the 1D configuration quickly saturates for increasing size $l$, while in the 2D configuration the slope sharply increases in both theory and experiment, as expected for a first order phase transition. }
    \label{fig:transition}
\end{figure*}
To investigate the steady-state behavior of the system and probe the phase transition, we solved Eq. (\ref{eq:gp}) numerically with the experimental parameters introduced in the previous section \footnote{Throughout the simulation results presented in this section, the cavity wedge is not taken into account for more efficient simulations. See supplementary material for more technical details on the simulation and the effect of the cavity wedge.} and the detuning is set to $\Dlp = \gamlp$ in the simulation (same value as in the experiments), which is in the regime where a driven-dissipative Kerr cavity exhibits mean-field bistability \cite{Baas2004Analogy, Rodriguez2017Probing}.
This can be equivalently viewed as the approximation of considering only the $k=0$ mode under uniform drive $F$ \cite{Casteels2017}, since the steady-state mean-field equation can be written as 
\bea\label{eq:MFbis}
\abs{\psiss}^2\left[ \left( \Dlp - \glp\abs{\psiss}^2 \right)^2 + \dfrac{\gamlp^2}{4} \right] & = \abs{F}^2.
\eea
Note that the non-linear relation between $\abs{\psiss}^2$ and $I=\abs{F}^2$ predicts a bi-stable regime if $\Dlp/\gamlp>\sqrt{3}/2$, as shown by the dashed line in Fig. \ref{fig:transition} (\new{d}), that we will compare with our numerical results. 
In all the simulations, the diameter of the probing disk $D$ is set to $l_D = 5\micron$ and in the 1D configuration the minor-axis of the driving spot is $b = 6.4\micron$, which are also the values adopted in our experiments.

In Fig. \ref{fig:transition} (a)-(c) [(\new{d})-(\new{f})] we present our theoretical and experimental results for the 1D (2D) driving geometry. In both configurations, the steady-state polariton density $\psibar$ averaged over the probing disk increases as a function of the driving intensity $I$ and the maximum slope $\mathcal{S}(l)=\max_I\{\frac{\partial \psibar(I,l)}{\partial I}\} $ of the crossover from low density to high density  (obtained with a noise-robust numerical differentiation method \cite{Chartrand2011Noise}) is monitored as a function of the top-hat size $l$, which allows us to probe the emergence of phase transitions defined by Eq. (\ref{eq:ordm}).
In the 1D configuration, where the top-hat drive takes the shape of an elliptical spot with fixed minor axis, the slope $\mathcal{S}(l)$ saturates to a finite value  with low enhancement [$\mathcal{S}(l)/\mathcal{S}(l_0)<2$ with $l_0=15\micron$ for all values of $l$ measured] as the major axis $l$ increases, signifying a smooth crossover with no phase transition in the thermodynamic limit.

In sharp contrast to the 1D configuration, with a 2D driving geometry, the slope presents a significant enhancement (by a factor of around 40 in theory, and a comparable value in the experimental results) as the top-hat diameter $l$ increases, suggesting the emergence of a first-order phase transition in the thermodynamic limit of $l\rightarrow\infty$. We would like to also point out that, while in the 1D configuration we observed no bistability, in the experiments with 2D geometry we observed slight bistability for top-hat diameters $l\gtrsim 35\micron$ [in this case we consistently took the lower branch when computing the slope (the higher one would give similar results)], which is consistent with the critical slowing down \footnote{See supplementary material for a discussion on the critical slowing down based on our numerical simulations.} of the dynamics  as the system approaches criticality in 2D.
Note that for $\mathcal{S}(l)/\mathcal{S}(l_0)\gtrsim 10$ [corresponding to a top-hat size of $l\gtrsim 30$ in the experimental results in Fig. \ref{fig:transition}(\new{e})], the curve becomes almost vertical [$\tan^{-1}(10)\simeq 84^\circ$], which makes the numerically computed derivatives more sensitive to small errors in the measurements, resulting in the relatively larger errorbars on the experimental curve in Fig. \ref{fig:transition}(\new{f}) in this regime \new{\footnote{\new{The deviation between the theoretical and the experimental curves could originate from a slightly higher detuning in the experiment than the nominal value $\Dlp=\gamlp$. Nevertheless, the main objective of this plot is to show the divergence in the 2D configuration for both theory and experiment, which is in clear contrast to the 1D case.}}}.

{\it Conclusion and outlook.---}
In this work, we have demonstrated both experimentally and theoretically the emergence of a first-order dissipative phase transition of polaritons in a planar microcavity subjected to a top-hat driving scheme with naturally diffusive boundary conditions. We have shown that the emergence of criticality in such photonic system with Kerr nonlinearity is determined by the spatial dimension via the geometry imposed by the top-hat driving spot: a 1D geometry leads to a crossover behavior with no phase transition, while a 2D geometry shows a behavior consistent with a first-order transition between two phases with different densities, which, to the best of our knowledge, is the first experimental demonstration of the role of dimensionality in determining criticality in driven-dissipative photonic systems.

The approach presented in this work allows the study of both 1D and 2D problems using the same planar cavity. The ability to control the criticality of the system via the spatial profile of the drive can also bring new insights to the design of polaritonic devices such as all-optical polariton transistors \cite{Ballarini2013}. This scheme can be potentially generalized to more complicated geometries imprinted by the shape of the driving field, such as fractal patterns or quasi-periodic lattices, which could open the possibilities for studying effects of gradual changes of the dimensionality on phase transitions, paving the way to a novel approach to exploring the many-body physics of photons and critical phenomena.

\acknowledgements{We would like to acknowledge discussions with Z. Denis. This work was supported by the FET FLAGSHIP Project PhoQuS (grant agreement ID: 820392) and by Project NOMOS (ANR-18-CE24-0026). This work was granted access to
the HPC resources of TGCC under the allocation
A0100512462 attributed by GENCI (Grand Equipement
National de Calcul Intensif). AB and QG are members of the Institut Universitaire de France (IUF).}

% \clearpage
\bibliography{bib}
\clearpage

\pagebreak
% \widetext

%%%%%%%%%% Merge with supplemental materials %%%%%%%%%%
\widetext
\clearpage
\begin{center}
\textbf{\large Supplemental Materials: \\Dissipative phase transition with driving-controlled spatial dimension \\ and diffusive boundary conditions \\}
\end{center}

% 
% \title{\large Supplemental Material: \\Dissipative phase transition with driving-controlled spatial dimension \\ and diffusive boundary conditions \\}

% \author{Zejian Li}
% \thanks{These two authors contributed equally}
% \affiliation{Universit\'{e} de Paris, Laboratoire Mat\'{e}riaux et Ph\'{e}nom\`{e}nes Quantiques, CNRS-UMR7162, 75013 Paris, France}

% \author{Ferdinand Claude}
% \thanks{These two authors contributed equally}
% \affiliation{Sorbonne Universit\'{e}, Laboratoire Kastler Brossel, CNRS-UMR8552, 75005 Paris, France}

% \author{Thomas Boulier}
% \affiliation{Sorbonne Universit\'{e}, Laboratoire Kastler Brossel, CNRS-UMR8552, 75005 Paris, France}

% \author{Elisabeth Giacobino}
% \affiliation{Sorbonne Universit\'{e}, Laboratoire Kastler Brossel, CNRS-UMR8552, 75005 Paris, France}

% \author{Quentin Glorieux}
% \affiliation{Sorbonne Universit\'{e}, Laboratoire Kastler Brossel, CNRS-UMR8552, 75005 Paris, France}

% \author{Alberto Bramati}
% \affiliation{Sorbonne Universit\'{e}, Laboratoire Kastler Brossel, CNRS-UMR8552, 75005 Paris, France}

% \author{Cristiano Ciuti}
% \affiliation{Universit\'{e} de Paris, Laboratoire Mat\'{e}riaux et Ph\'{e}nom\`{e}nes Quantiques, CNRS-UMR7162, 75013 Paris, France}

% \date{\today}
% \maketitle
\onecolumngrid

%%%%%%%%%% Merge with supplemental materials %%%%%%%%%%
%%%%%%%%%% Prefix a "S" to all equations, figures, tables and reset the counter %%%%%%%%%%
\setcounter{equation}{0}
\setcounter{figure}{0}
\setcounter{table}{0}
\setcounter{page}{1}
\makeatletter
\renewcommand{\theequation}{S\arabic{equation}}
\renewcommand{\thefigure}{S\arabic{figure}}
% \renewcommand{\bibnumfmt}[1]{[S#1]}
% \renewcommand{\citenumfont}[1]{S#1}
%%%%%%%%%% Prefix a "S" to all equations, figures, tables and reset the counter %%%%%%%%%%

\section{Fully quantum description of the system and derivation of the mean-field dynamics}\label{app:quant}

The lower polariton field can be represented by the quantum operator $\hat{\psilp}(\vec{r},t)$, where the field operators obey the standard bosonic commutation relations 
% \bea
$[\hat{\psi}(\vec{r}),\hat{\psi}(\vec{r}')]=0$, $ [\hat{\psi}(\vec{r}),\hat{\psi}^\dagger(\vec{r}')]=\delta(\vec{r}-\vec{r}')$.
% \eea
In the frame rotating at the driving frequency $\omega_d$, the Hamiltonian of the system can be written as \cite{Carusotto2013Quantum}
\bea
\hat{\mathcal{H}}=& \int \d^2 \vec{r}~ \hat{\psi}^\dagger(\vec{r})\left[ -\hbar\Dlp\hat{\psi}(\vec{r}) - \dfrac{\hbar^2\nabla^2}{2\mlp}\hat{\psi}(\vec{r}) \right]\\
&+ \int\d^2\vec{r}~\dfrac{\hbar\glp}{2}\hat{\psi}^{\dagger 2}(\vec{r})\hat{\psi}^2(\vec{r})\\
&+ \int\d^2\vec{r}~\hbar\left[ \fcal(\vec{r})\hat{\psi}^\dagger(\vec{r}) + \fcal^*(\vec{r})\hat{\psi}(\vec{r}) \right].
\eea
Under the Born-Markov approximation and assuming a uniform single-polariton loss rate $\gamlp$, the dynamics of the field can be cast in a Lindblad form in terms of its density matrix $\rhohat$ \cite{Breuer2007}:
\bea\label{eq:lind}
\dfrac{\d \rhohat}{\d t} =& \lcal[\rhohat]= -\dfrac{\rmi}{\hbar}\left[ \hat{\mathcal{H}},\rhohat \right] + \mathcal{D}\left[\rhohat\right],
\eea
with the dissipator
\bea
\mathcal{D}\left[\rhohat\right]=& \int\d^2\vec{r}~\dfrac{\gamlp}{2}\left[ 2\hat{\psi}(\vec{r})\rhohat\hat{\psi}^\dagger(\vec{r}) -\left\{ \hat{\psi}^\dagger(\vec{r})\hat{\psi}(\vec{r}),\rhohat\right\}\right].
\eea
Within the mean-field approximation, where one assumes the factorization of the expectation value \bea\label{eq:mfassum}
\langle \hat{\psi}^{\dagger m}(\vec{r})\hat{\psi}^n(\vec{r}) \rangle\simeq\langle\hat{\psi}^{\dagger}(\vec{r})\rangle^m\langle\hat{\psi}(\vec{r})\rangle^n,
\eea
the dynamics of mean-field parameter
$\psilp(\vec{r},t)=\langle \hat{\psi}(\vec{r},t) \rangle$ can be derived as 
\bea
    \dfrac{\partial \psi}{\partial t} =&  \dfrac{\partial}{\partial t}\tr{[\rhohat\hat{\psi}]} = \tr{\left[ \dfrac{\d \rhohat}{\d t} \hat{\psi} \right]} ,
\eea
which, after straightforward calculations using Eq. (\ref{eq:lind}) and (\ref{eq:mfassum}) , gives the dynamical equation of $\psi$ introduced in the main text.

\section{Driven-dissipative dynamics in polar coordinates}\label{ap:polar}
In the configuration where a 2D round top-hat is applied, we can efficiently simulate the mean-field equation by adopting the polar coordinates and taking advantage of the cylindrical symmetry of the problem.
The Laplacian of a scalar field $\phi$ in polar coordinates $(r,\theta)$ can be written as
\bea
\nabla^2 \phi &= \partial_r^2\phi+\dfrac{1}{r^2}\left( \partial_\theta^2\phi+r\partial_r\phi\right).
\eea
In the presence of cylindrical symmetry (where we ignore the effect of the wedge), we can look for solutions of the form $\psilp=\psilp(r,t)$ that have no angular dependence, which means $\partial_\theta\psilp=0$. Therefore, Eq. (1) in the main text becomes the radial equation
\bea
\rmi \dfrac{\partial}{\partial t}\psilp(r,t) =& \left[ -\Dlp - \dfrac{\hbar}{2\mlp}\left( \partial_r^2+\dfrac{1}{r}\partial_r \right) \right]\psilp \\&+ \glp\abs{\psilp}^2\psilp - \rmi\dfrac{\gamlp}{2}\psilp \\&+ \fcal(r),
\eea
which significantly reduces the computational cost.

\section{Benchmark of the mean-field equation against the truncated Wigner approximation method}\label{app:tw}

To justify the use of the mean-field (MF) equation for the numerical results presented in this article, in this section, we benchmark the solutions against the truncated Wigner (TW) approximation method \cite{Vogel1989,Carusotto2013Quantum}, which requires discretizing the quantum lower-polariton field $\hat{\psi}$ in the cavity into a 2-dimensional lattice. We denote $\varphi_j=\varphi(\vec{r}_j)$ the complex field amplitude (not to be confused with the mean-field classical parameter $\psi$ used in the main text) at the lattice position $\vec{r}_j$, and $\Delta V=\Delta x^2$ the size of the elementary cell of the discrete lattice (where we adopted a step length of $\Delta x$ when discretizing both dimensions). Note that the discretized field operators satisfy the commutation relation $ [\hat{\psi}_j,\hat{\psi}^\dagger_{j'}]=\delta{j,j'}/\Delta V $. The time evolution of the discretized field can be exactly described by a third-order partial differential equation in terms of its Wigner distribution (which is a representation of the density matrix). In the limit of $ \glp/(\gamlp \Delta V)\ll 1 $, the third-order derivative terms can be neglected, resulting in a Fokker-Planck equation \cite{Carmichael1999Statistical}, that can be solved using stochastic trajectories defined via the set of Langevin equations in terms of the amplitude $\varphi_j$:

\bea
\d \varphi_j(t) =& F_j\{\varphi\}\d t+\sqrt{\dfrac{\gamlp}{4\Delta V}}\d W_j,
\eea

where $F_j\{\varphi\}=F\{\varphi\}(\vec{r}=\vec{r}_j)$ is the drift force on the lattice site $j$, with
\bea
    F\{\varphi\}(\vec{r}) =& -\rmi \left[ -\Dlp - \dfrac{\hbar}{2\mlp}\nabla^2 -\rmi\dfrac{\gamlp}{2} \right.\\
     &+\left. \glp\left( \abs{\varphi(\vec{r},t)}^2-\dfrac{1}{\Delta V} \right)\right]\varphi(\vec{r},t)\\
     &+ \fcal(\vec{r})
\eea
and $\d W_j$ is a zero-mean complex Gaussian noise satisfying
\bea
    \d W_j\d W_{j'} =& 0,\\
    \d W^*_j\d W_{j'}=&2\d t\delta_{j,j'}.
\eea
Within this formalism, the expectation values for symmetrized products of field operators \cite{Vogel1989} are given by the statistical expectation over different stochastic trajectories:
\bea
    \left\langle \{ ( \hat{\psi}_j^\dagger)^n,\hat{\psi}_k^m ) \}_s\right\rangle =& \mathbb{E}\left[ (\varphi_{j,r}^*)^n\varphi_{k,r}^m \right]\\\simeq& \dfrac{1}{N_\mathrm{traj}}\sum_r (\varphi_{j,r}^*)^n\varphi_{k,r}^m,
\eea
where the index $r$ labels the $N_\mathrm{traj}$ random trajectories. For example, the polariton density at each site, that we are interested in, can be calculated as 
\bea
n_j=\left\langle \hat{\psi}_j^\dagger\hat{\psi}_j \right\rangle =& \left\langle \dfrac{1}{2}( \hat{\psi}_j^\dagger\hat{\psi}_j + \hat{\psi}_j\hat{\psi}^\dagger_j ) - \dfrac{1}{2}[\hat{\psi}_j,\hat{\psi}^\dagger_{j}] \right\rangle \\
=& \left\langle \{ \hat{\psi}^\dagger_j,\hat{\psi}_j \}_s \right\rangle - \dfrac{1}{2\Delta V}\\
\simeq& \dfrac{1}{N_\mathrm{traj}}\sum_r \varphi_{j,r}^*\varphi_{j,r} - \dfrac{1}{2\Delta V}.
\eea

\begin{figure}[h]
    \centering
    \includegraphics[width=0.65\linewidth]{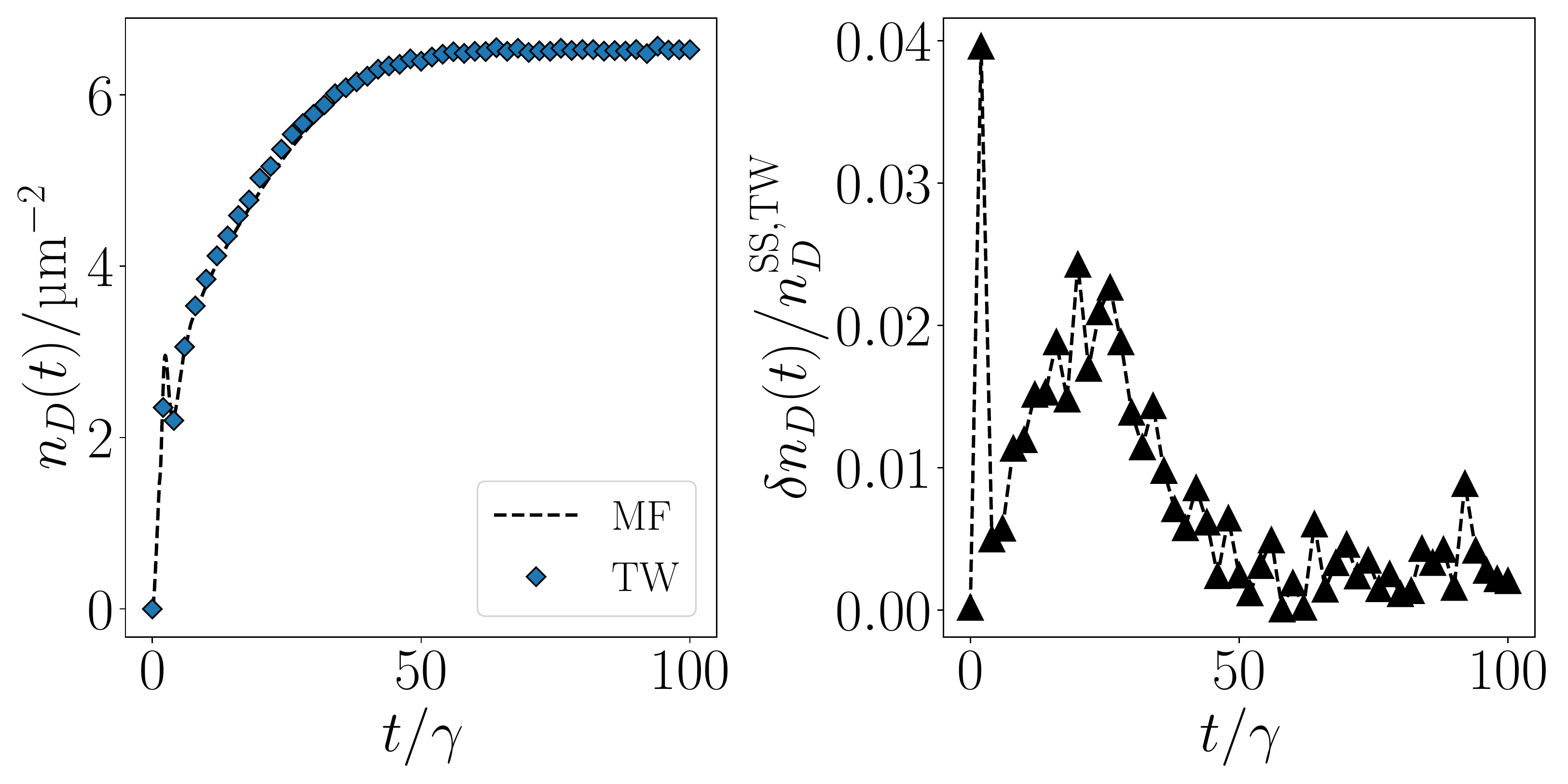}
    \caption{Left: time evolution of the polariton density averaged over the probing disk, computed with both MF (dashed line) and TW (diamonds) methods. Right: relative error in the MF results with respect to the steady state density, as a function of time, where $\delta n_D(t)=\lvert n^{\mathrm{TW}}_D(t)-n^{\mathrm{MF}}_D(t)\rvert$. The driving is $F = 1.35\gamma/\micron$, $(I = \abs{F}^2 = 1.8225\gamma^2/\micron^2)$ with a top-hat size of $l=45\micron$. Error bars are within the symbol size.}
    \label{fig:bm_nt}
\end{figure}

\begin{figure}[h]
    \centering
    \includegraphics[width=0.65\linewidth]{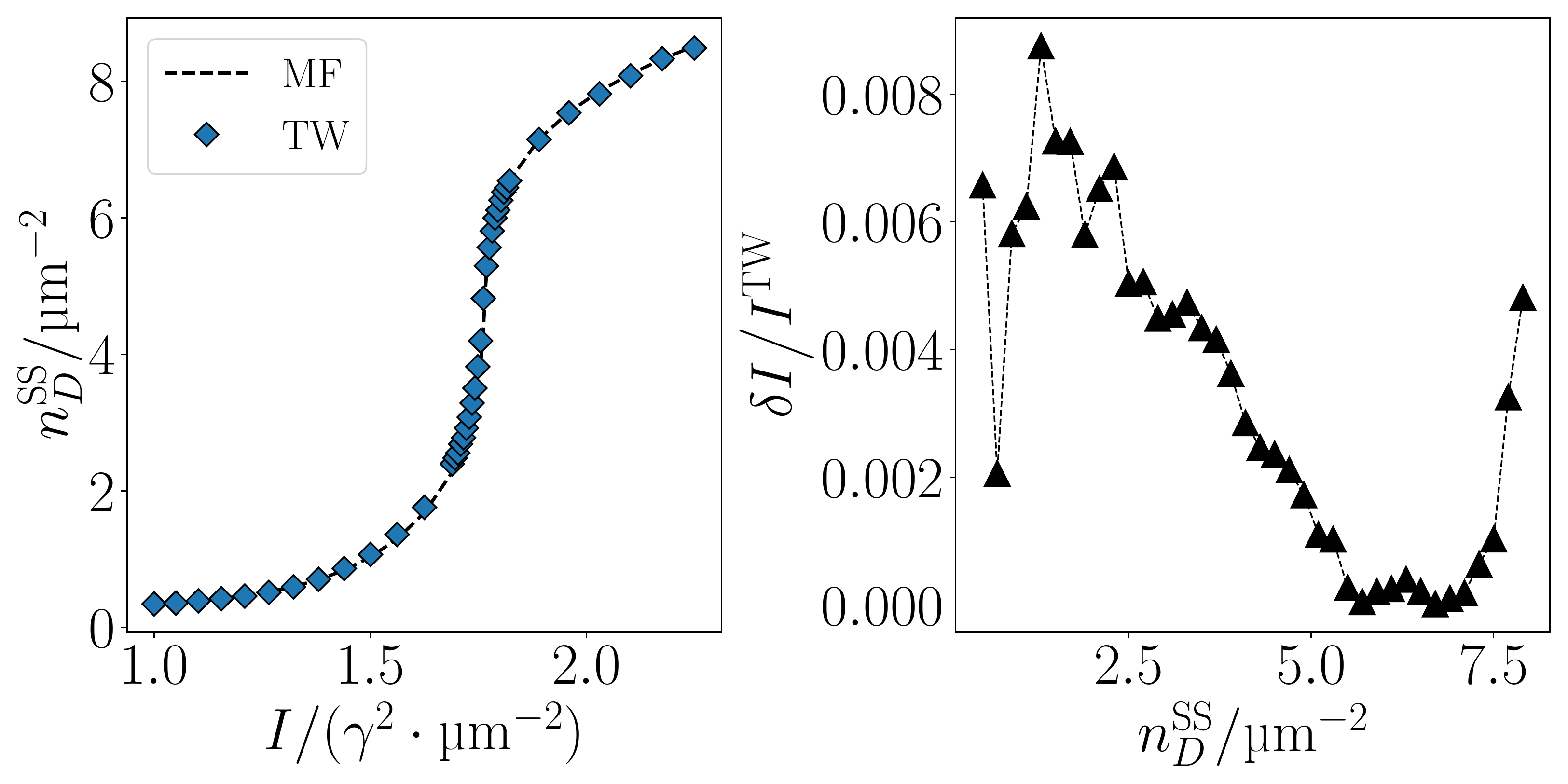}
    \caption{Left: steady-state polariton density averaged over the probing disk as a function of the drive, computed with both MF (dashed line) and TW (diamonds) methods. Right: relative error of the driving in the MF results as a function of the density, where $\delta I =\left\lvert I^\mathrm{TW}-I^\mathrm{MF}\right\rvert$. The top-hat size is $l=45\micron$. Error bars are within the symbol size.}
    \label{fig:bm_nss}
\end{figure}

We performed the truncated Wigner simulations for the 2D configuration with a top-hat spot $l=45\micron$ driving the centre of a $105\micron\times105\micron$ square lattice (this corresponds to a non-driven region with minimum width $30\micron$ surrounding the driving spot, which, as we verified, is sufficient for the result to converge), where the discretization is set to $\Delta x = 2\micron$ (thus $\glp/(\gamlp\Delta V) = 0.03125$). Note that the lattice has $53\times53=2809$ sites, which is the number of coupled stochastic differential equations to solve in each single trajectory. For comparison, we simulate the same configuration ($l=45\micron$ at the centre of a disk with diameter $105\micron$) with the MF equation in polar coordinates as introduced in Section \ref{ap:polar}, with a discretization step length of $\Delta r = 0.5\micron$ in the radial dimension, where we have only $105$ equations to solve. In Fig. \ref{fig:bm_nt} we compare the time evolution of the polariton density averaged over the probing disk with driving $F = 1.35\gamma/\micron$, $(I = \abs{F}^2 = 1.8225\gamma^2/\micron^2)$, simulated with the two aforementioned methods, where we averaged over $N_\mathrm{traj}=1000$ trajectories in the TW simulation. The relative error in the polariton density stays well below $5\%$ of the steady-state density throughout the time evolution, and decreases to less than $1\%$ as the steady state is reached, showing a good agreement between the two. In Fig. \ref{fig:bm_nss} we compare the steady-state polariton density averaged over the probing disk for different driving intensities across the crossover, and we took $N_\mathrm{traj}=100$ for each driving value due to the high computational cost. The relative error in the driving intensity stayed well below $1\%$ throughout the crossover. We therefore choose to use the mean-field equation for the numerical study presented in the main text, which gives accurate results at much lower computational cost as compared to the truncated Wigner method.

\section{Critical slowing down of the dynamics in the 2D configuration}

\begin{figure}
    \centering
    \includegraphics[width=0.5\linewidth]{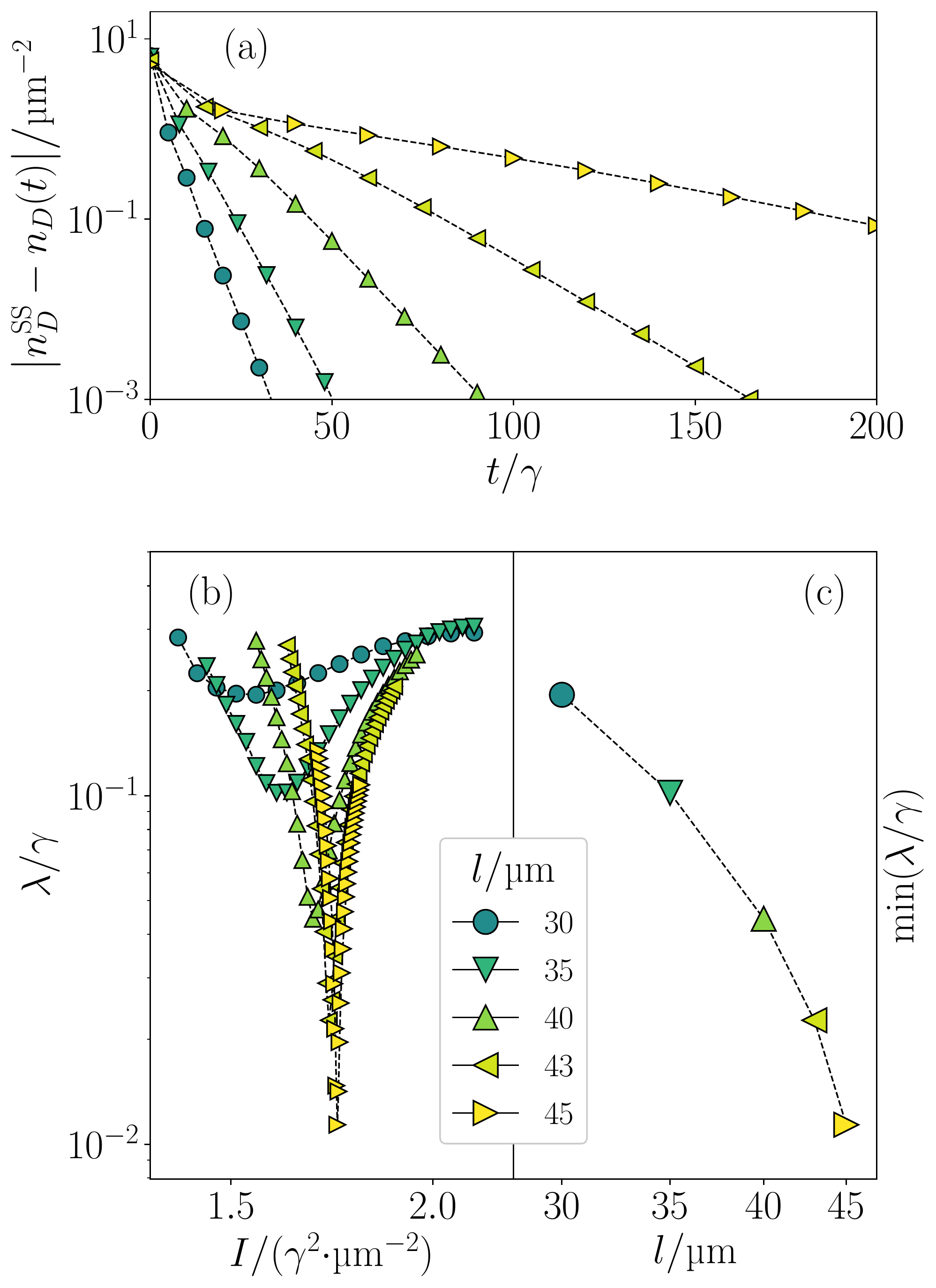}
    \caption{Critical slowing down and the closing of the Liouvillian gap for a 2D top-hat drive observed in numerical simulations. (a) The relaxation of $\psid(t)$ towards the steady state $\psibar$ for difference driving spot sizes $l$ (see legend) at $I = 1.7689\gamlp^2/\micron^2$. The other parameters are the same as in the main text. (b) The Liouvillian gap $\lambda$ evaluated from the asymptotic decay rate as a function of the driving intensity $I$ for different values of $l$. The error bars are within the symbol size. (c) The minimum of $\lambda$ as a function of $l$.  }
    \label{fig:liou}
\end{figure}

The Liouvillian superoperator $\lcal$ [as introduced in Eq. (\ref{eq:lind})] has eigenvalues $\left\{ \lambda_j ~ | ~ \exists \rhohat_j, \lcal[\rhohat_j] = \lambda_j\rhohat_j \right\}$ with $\Re(\lambda_j)\leq 0$. We call $\lambda = \min \left\{ \left\lvert \Re(\lambda_j)\right\rvert \right\}$ the Liouvillian gap, which governs the slowest relaxation dynamics of the density matrix towards the steady state and which is the inverse of the asymptotic decay rate \cite{Minganti2018Spectral}. A dissipative phase transition can be characterised by the closing of the Liouvillian gap $\lambda\rightarrow 0$ in the thermodynamic limit, and is associated with a critical slowing down in the transient dynamics \cite{Vicentini2018Critical,Minganti2018Spectral}.

To further investigate the dynamical properties of the emerging criticality in 2D, we simulate the time evolution of the polariton density $\psid(t)$ averaged over the probing disk towards the steady state $\psibar$ in the 2D configuration, with a vacuum initial state. For driving intensities close to the critical point, the difference $\psid(t) - \psibar$ decays exponentially to zero for large time scales $t\gg 1/\gamlp$, as reported in Fig. \ref{fig:liou} (a) showing the particular case of $I = 1.7689\gamlp^2/\micron^2$ for different driving spot sizes $l$. The decay exhibits a critical slowing down as $l$ increases, and we can estimate the Liouvillian gap $\lambda$ by fitting the decay to an exponential form $\psid(t) = \psibar + A \exp(-\lambda t)$, as the asymptotic decay is dominated by the Liouvillian gap in this regime \cite{Vicentini2018Critical,Minganti2018Spectral}. The dependence of $\lambda$ as a function of the driving intensity $I$ and spot size $l$ is quantified in Fig. \ref{fig:liou} (b) and (c): $\lambda(I)$ presents a dip for each size $l$, and the minimum keeps decreasing as the driven area is increased. This evidence suggests the closing of the Liouvillian gap $\lambda$ in the thermodynamic limit, confirming the emergence of the first-order dissipative phase transition from the dynamical point of view. We can also estimate the critical driving intensity to be $I_c\simeq 1.76\gamlp^2/\micron^2$ from this figure.

Note that in our theoretical results, unlike the prediction of the single-mode mean-field theory given by Eq.(7) in the main text, where the steady-state should always exhibit bistability across the transition, no bistability has been observed throughout the simulation results presented in the main text (for larger sizes in 2D we would expect its appearance as we approach the limit of driving only the $k=0$ mode), despite the absence of quantum fluctuations in the mean-field equation. This can be explained by the fact that our top-hat drive (that is non-uniform across the planar microcavity) excites both the $k=0$ mode and other $k\neq 0$ modes. While only the $k=0$ mode is responsible for the phase transition \cite{Casteels2017}, the other $k\neq 0$ modes serve as a reservoir in the Fourier $k$ space, whose fluctuations, despite being at the mean-field level, suffice to suppress the bistability.

\section{Effect of the cavity wedge}\label{ap:wedge}

\begin{figure}[htbp]
    \centering
    \includegraphics[width=0.5\linewidth]{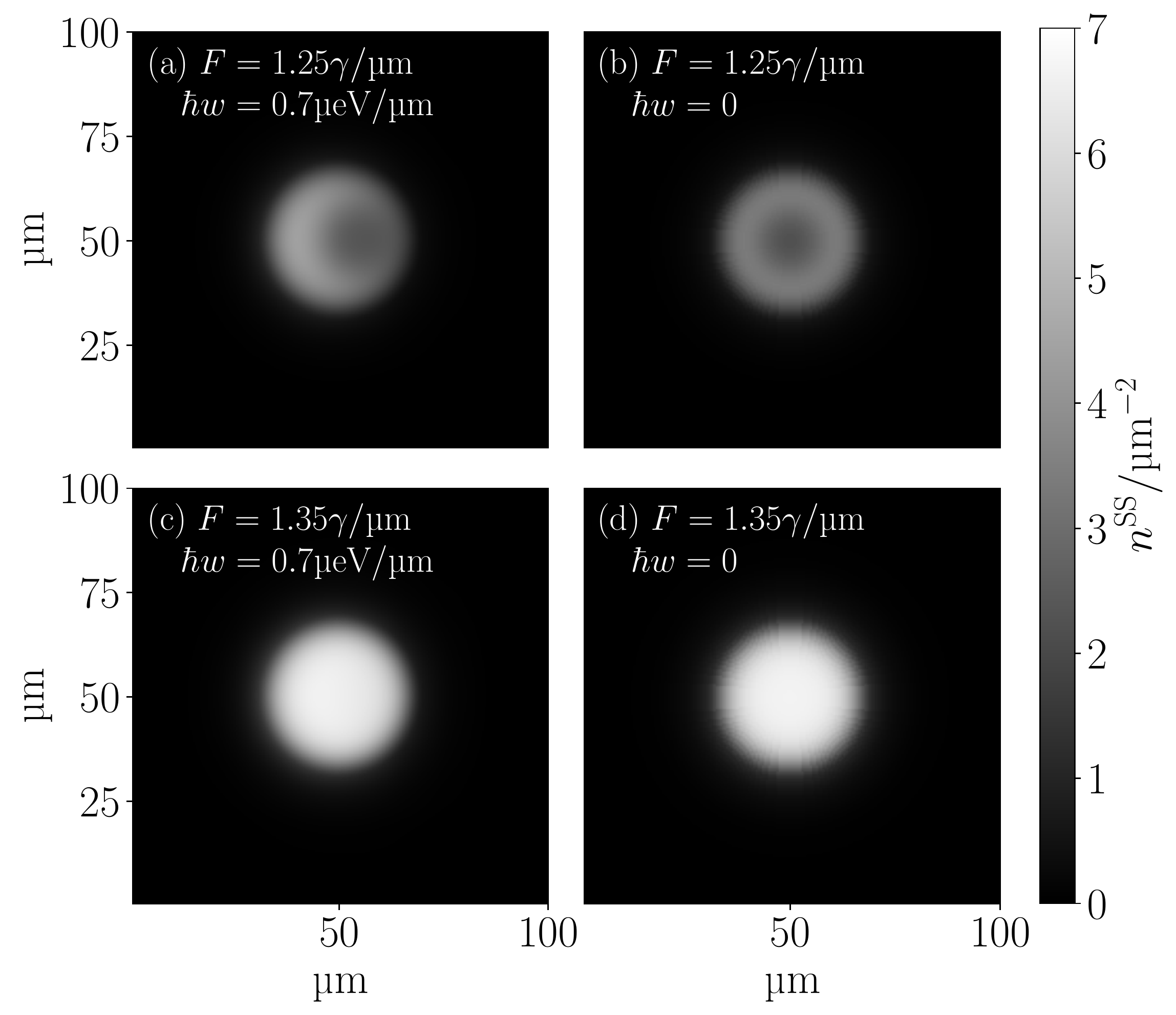}
    \caption{Steady-state polariton density (see colorbar) distribution in the planar microcavity simulated with experimental parameters with a 2D top-hat ($l=40\micron$) for different driving and wedge values (see annotation); the detuning with respect to the center of the driving spot is $\Dlp=\gamlp$. The left (resp. right) column corresponds to the wedge value $\hbar w = 0.7$\textmu eV (resp. $w = 0$). The driving in the top (resp. bottom) row corresponds to the lower (resp. higher) density phase close to the crossover. }
    \label{fig:wedgeHeatMaps}
\end{figure}

\begin{figure}[htbp]
    \centering
    \includegraphics[width=0.7\linewidth]{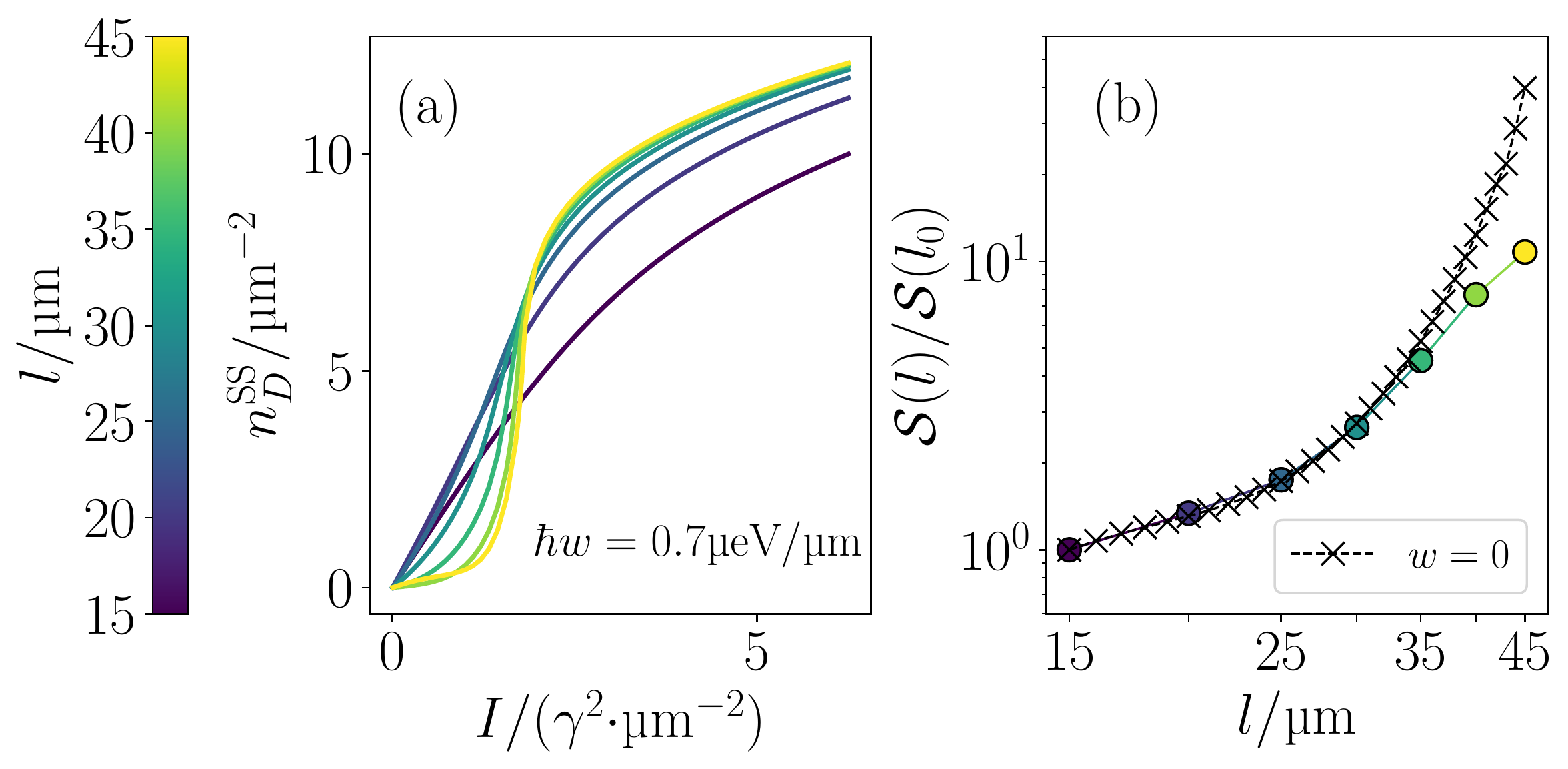}
    \caption{(a) Steady-state polariton density $\psibar$ averaged over the probing disk as a function of driving intensity $I$ simulated with the experimental parameters for different 2D top-hat spot sizes $l$ (see colorbar) with the wedge $\hbar w = 0.7$ µeV/µm taken into account, and the detuning at the center of the top-hat is $\Dlp = \gamlp$. (b) The maximum crossover slope of each top-hat size $\mathcal{S}(l)$ normalized by the maximum slope at $l_0=15\micron$ in log-log scale, and the dashed line with ``x" markers represents the results with $w=0$ in the simulation as presented in the main text.}
    \label{fig:wedge}
\end{figure}

In order to study the effect of the cavity wedge that we excluded in the simulation presented in the main text for a more efficient simulation of the mean-field equation, we present in this section the simulation results of the 2D configuration with the wedge taken into account. The detuning $\Dlp$ becomes position dependent and can be modelled as
\bea
\Dlp(x,y)=& \Dlp(x_0) + w(x-x_0),   
\eea
where we have orientated the $x$ axis along the gradient of the wedge. The steady-state polariton density distribution of the lower  phase in the planar microcavity typically exhibits a distorted ring-shaped pattern, shown in Fig. \ref{fig:wedgeHeatMaps} (a) and (b), as the result of the wedge breaking the cylindrical symmetry of the system, whereas the distorsion is less visible in the higher density phase, as shown in Fig. \ref{fig:wedgeHeatMaps} (c) and (d). Despite the distortion, we obtain qualitatively similar behavior of the steady state as a function of the drive (Fig. \ref{fig:wedge}) compared to the case without wedge (as presented in the main text). Note that for top-hat sizes $l\lesssim 40\micron$ (which is the main range of our experimental results), the difference in the maximum slope between the two cases remains minor, which justifies our choice of neglecting the wedge in 2D simulations when comparing the results to the experiments as done in the main text, in return for more efficient simulations. On the other hand, for larger driving spots in 2D experiments where the wedge is present, we should expect a less significant growth or even the saturation of the slope compared to the ideal wedgeless scenario.

\end{document}